\documentclass[%
 reprint,
 superscriptaddress,
preprintnumbers,
nofootinbib,
 amsmath,amssymb,
 aps,
prl,
]{revtex4-1}

\usepackage{amsmath}
\usepackage{amssymb}
\usepackage{amsthm}
\usepackage{MnSymbol}
\usepackage{nicefrac}
\usepackage{amsfonts}
\usepackage{dsfont}

\usepackage[markup=nocolor]{changes}
\usepackage{natbib}

\usepackage{graphicx}

\usepackage[normalem]{ulem}

\usepackage{dcolumn}
\usepackage{bm}

\usepackage{color}
\newcommand{\bea}{\begin{eqnarray}}
\newcommand{\eea}{\end{eqnarray}}
\newcommand{\be}{\begin{equation}}
\newcommand{\ee}{\end{equation}}

\begin{document}

\title{Suppression of proton decay in quantum gravity}
 
 \author{Astrid Eichhorn}
   \email{eichhorn@cp3.sdu.dk}
\affiliation{CP3-Origins, University of Southern Denmark, Campusvej 55, 5230 Odense M, Denmark}
\author{Shouryya Ray}
\email{sray@cp3.sdu.dk}
\affiliation{CP3-Origins, University of Southern Denmark, Campusvej 55, 5230 Odense M, Denmark}
\affiliation{Institut f\"ur Theoretische Physik and W\"urzburg-Dresden Cluster of Excellence ct.qmat, TU Dresden, 01062 Dresden, Germany}

\begin{abstract}
The bound on the proton lifetime is one of a small handful of observations that constrains physics not far from the Planck scale. This calls for a calculation of the proton lifetime in quantum gravity. Here, we calculate how quantum fluctuations of the metric impact four-fermion interactions which mediate proton decay. We find that quantum gravity lowers the scaling dimension of the four-fermion interaction, suppressing proton decay and raising the proton lifetime significantly.

As a special case, we analyze asymptotically safe quantum gravity, in which we discover that proton-decay-mediating four-fermion interactions are strictly zero.
\end{abstract}


\maketitle

\emph{Introduction: the proton lifetime constrains quantum gravity.}\\
The proton is an exceptionally long-lived triquark bound state: the current lower bound for its lifetime has been estimated at Super-Kamiokande to be of the order of $10^{34}$ years \cite{ParticleDataGroup:2022pth}, with Hyper-Kamiokande, DUNE and JUNO projected to improve the bound by about an order of magnitude \cite{Hyper-Kamiokande:2018ofw,DUNE:2016hlj,JUNO:2015zny}. This is quite a stringent lower bound and among only a handful of measurements that constrain theories of matter and fundamental interactions at scales of $10^{16}\, \rm GeV$ and above. The constraint arises as follows: Theoretically, the proton lifetime $\tau_p$ is related to the mass $M_{X}$ of a hypothetical particle that mediates proton decay. In terms of $M_X$ and the proton mass $M_p$, the proton lifetime is estimated as
\be
  \tau_p 
  \approx
  M_p^{-1} \left( M_X / M_p \right)^4,
    \label{eq:taup-master}
\ee
The observational constraint $\tau_p> 10^{34}$ years thus translates into $M_X \gtrsim 2\cdot 10^{16}\, \rm GeV = 2\,M_{\rm GUT}$. That this is a stringent constraint on grand unified theories (GUTs) is well-explored, see, e.g.,  \cite{Langacker:1980js,Nath:2006ut,King:2021gmj}. %
However, GUTs are not the only new physics that may exist at or above $M_{\rm GUT}$; the quantum-gravity scale is also nearby. How closely nearby it is, depends on the quantum-gravity theory: the reduced Planck mass lies at $10^{18}\, \rm GeV$; the actual quantum-gravity scale may be even lower (or higher) than that. For instance, in settings with sufficiently many matter fields, the quantum-gravity scale may be as low as the GUT scale \cite{Calmet:2011ic,deAlwis:2019aud}, in settings with extra dimensions, it may even be much lower \cite{Antoniadis:1998ig}, implying that the bound on the proton lifetime constrains these settings \cite{Adams:2000za}. There are also some discrete quantum-gravity approaches in which effects of quantum gravity are expected to set in far below the Planck scale, e.g., \cite{Belenchia:2014fda}. Finally, in the context of black-hole physics, it is conjectured that effects of quantum gravity are relevant at the horizon, implying that the scale of quantum gravity may be significantly lower than the Planck scale, see, e.g., \cite{Almheiri:2012rt,Cardoso:2016oxy,Haggard:2016ibp,Eichhorn:2022bgu} and references therein.

\emph{Quantum gravity and proton decay through four-fermion couplings.}\\
This clearly calls for the calculation of the proton decay rate in quantum gravity, such that the experimental constraint $\tau_p>10^{34}\, \rm years$ can be brought to bear on theory-development in quantum gravity. Quantum gravity theories come with diverse assumptions about the degrees of freedom, the dynamics and even the mathematical formalism, see, e.g., \cite{deBoer:2022zka}. They all have in common that inbetween the full quantum gravity theory in the ultraviolet (UV) and General Relativity in the infrared (IR), there is a regime in which gravity can be quantized as an effective field theory \cite{Donoghue:1994dn,Burgess:2003jk,Donoghue:2022eay}. In this regime, the quantum fluctuations of gravity are quantum fluctuations of the metric, described within the standard local quantum field theory framework. We calculate the effect of quantum gravity on the proton lifetime within this regime. To do so, we extend the Standard Model effective field theory (SMEFT), see \cite{Willenbrock:2014bja,Brivio:2017vri,Falkowski:2015fla} for reviews, into the quantum-gravity regime, where we add metric fluctuations to the set of quantum fluctuations of the SM fields.

The SMEFT contains all interactions compatible with the SM symmetries. This includes four-fermion interactions for proton decay with corresponding coupling $\bar{G}_{4\rm F}^{qqq\ell}$, see, e.g., \cite{Grzadkowski:2010es,Falkowski:2017pss}. The proton lifetime is inversely proportional to the square of $G_{4\rm F}^{qqq\ell}(M_p)$,
\be
\tau_p \approx 16 \pi M_p^{-1} \left(G_{4\rm F}^{qqq\ell}(M_p)\right)^{-2},
\ee
where
\begin{align}
G_{4\rm F}^{qqq\ell}(k) = \bar{G}_{4\rm F}^{qqq\ell}(k) k^2
\end{align}
is the dimensionless four-fermion interaction with $k$ the Renormalization Group (RG) scale, and we have included the explicit $16\pi$ phase-space factor for decay rates \cite{Manohar:2018aog} for four-fermion interactions.

In perturbation theory, the leading-order scaling of $G_{4\rm F}^{qqq\ell}(k)$ is quadratic, such that $\bar{G}_{4\rm F}^{qqq\ell}$ is in fact constant. As a function of the scale $M_X$, we thus have
\be
G_{4\rm F}^{qqq\ell}(k) = G_{4\rm F}^{qqq\ell}(M_X) \frac{k^2}{M_X^2}.
\ee
To calculate the proton lifetime, we set $k = M_p$ to obtain 
\be
\tau_p \approx 16\pi M_p^{-1} \left( G_{4\rm F}^{qqq\ell}(M_X)\right)^{-2} \frac{M_X^4}{M_p^4}.\label{eq:protontauQGcanonical}
\ee
If we insert $M_X = M_{\rm GUT} = 10^{16}\, \rm GeV$, we obtain $\tau_p \lesssim \left( G_{4\rm F}^{qqq\ell}(M_X)\right)^{-2} \cdot 10^{32}$ years. With $G_{4\rm F}^{qqq\ell}(M_X) \sim \mathcal{O}(1)$, $M_X= M_{\rm GUT}$ is thereby already ruled out.

\begin{figure}[!t]
\includegraphics[width=\linewidth, clip=true, trim=0cm 22cm 0cm 0cm]{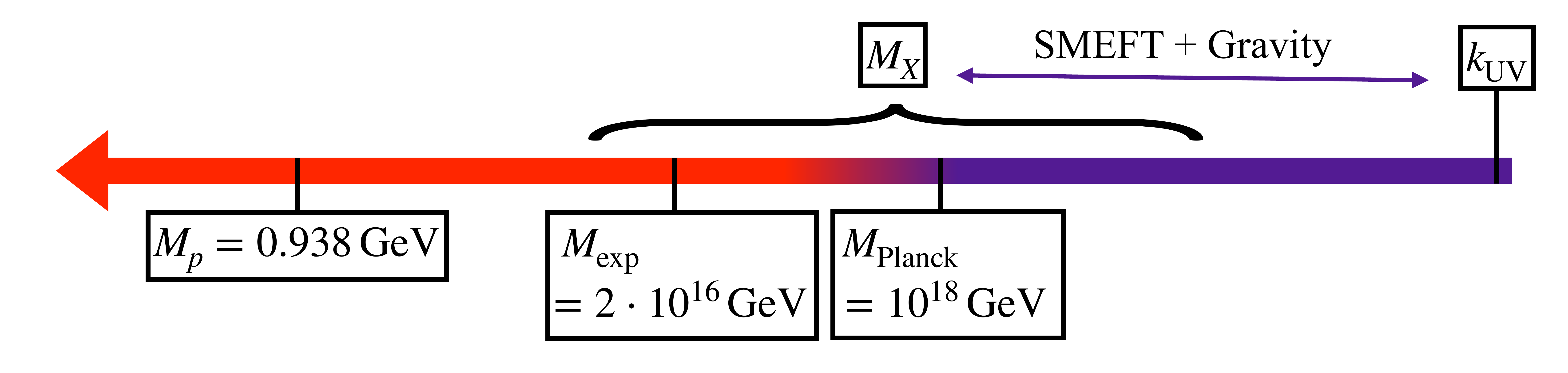}
\caption{\label{fig:illuscales} We illustrates the scales in our setting: In the far UV, at $k_{\rm UV}$, a quantum-field theoretic description of the Standard Model plus gravity breaks down; some unknown quantum theory of gravity takes over. Inbetween $k_{\rm UV}$ and $M_X$, the Standard Model plus gravity can be described as an effective field theory and our calculation holds. The reduced Planck mass $M_{\rm Planck} = 10^{18}\, \rm GeV$ is expected to lie in the range covered by $M_X$, as is the scale $M_{\rm exp} = 2\cdot 10^{16}\, \rm GeV$, which is already constrained by proton lifetime bounds. In the deep IR, we find the proton mass $M_p=0.938\, \rm GeV$.}
\end{figure}

To determine whether or not the assumption $G_{4\rm F}^{qqq\ell}(M_X) \sim \mathcal{O}(1)$ is justified, we need to calculate the size of the four-fermion coupling.  We thus evaluate the effect of metric fluctuations on the scaling dimension of $G_{4\rm F}^{qqq\ell}(k)$. We use functional RG techniques \cite{Wetterich:1992yh,Morris:1993qb,Reuter:1996cp}, see \cite{Dupuis:2020fhh} for a review. These techniques allow us to integrate out quantum fluctuations inbetween a UV scale $k_{\rm UV}$ and the scale $M_X$, see Fig.~\ref{fig:illuscales} for an illustration of the scales involved in our setting. In this setup, the propagator for metric fluctuations is derived from the Einstein-Hilbert action, and thus parametrized by the Newton coupling $G_N$ and the cosmological constant $\Lambda_{\rm cc}$; see the supplementary material for further details.

We obtain
\be
\beta_{G_{4\rm F}^{qqq\ell}} = k\partial_k\, G_{4\rm F}^{qqq\ell}(k) = \left(2+ \eta_{4\rm F} \right)G_{4\rm F}^{qqq\ell} + \mathcal{O}(G_{4\rm F}^{qqq\ell})^2,\label{eq:betafunction}
\ee
with the quantum-gravity contribution
\bea
    \eta_{4\text{F}} &=& 2g_\text{N} \Biggl[ \frac{30}{24\pi (1 - 2\lambda_{\text{cc}})^2} 
    + \frac{6}{4\pi (- 3 + 4\lambda_{\text{cc}})^2} \nonumber\\
    &{}&- \frac{ 6(9 - 4\lambda_\text{cc})}{5\pi(-3 + 4\lambda_\text{cc})^2} 
  + \frac{9(3-2\lambda_\text{cc})}{4\pi (- 3 + 4\lambda_\text{cc})^2} \Biggr].\label{eq:eta4F}
\eea
It depends on the dimensionless Newton coupling $g_{\rm N} = G_N \cdot k^2$ and the dimensionless cosmological constant $\lambda_{\rm cc} = \Lambda_{\rm CC} \cdot k^{-2}$. Assuming that the cosmological constant is sufficiently small at all scales, we can expand the quantum-gravity induced scaling dimension to obtain
\be
\eta_{4\rm F} = \frac{29}{15\pi} g_{\rm N} + \frac{
{77}
}{9 \pi} g_{\rm N}\, \lambda_{\rm cc} + \mathcal{O}(\lambda_{\rm cc}^2).\label{eq:eta4Fexp}
\ee
Our central result is a positive gravitational contribution to the scaling dimension of the four-fermion operator, i.e., the four-fermion coupling is rendered more irrelevant by quantum gravity than it is canonically. {This holds for $\lambda_{\rm cc}>-9.05$, which is expected to be the phenomenologically relevant region.}
The dimensionless product $\Lambda = g_{\rm N}\cdot \lambda_{\rm cc}$ also contributes positively to $\eta_{4\rm F}$ in regions where $\lambda_{\rm cc} > 0$; such a dimensionless contribution from quantum gravity is expected to be universal. Eq.~\eqref{eq:eta4F} is also equal to the quantum-gravity contribution to other four-fermion operators, first calculated in \cite{Eichhorn:2011pc}. This equality is a consequence of the ``flavor-blindness" of quantum gravity: metric fluctuations are insensitive to the type of SM fermion they couple to. Based on this argument, the agreement between our result and \cite{Eichhorn:2011pc} is expected; that it comes out of the explicit calculation is a nontrivial cross-check.

\begin{figure}
\includegraphics[width=\linewidth]{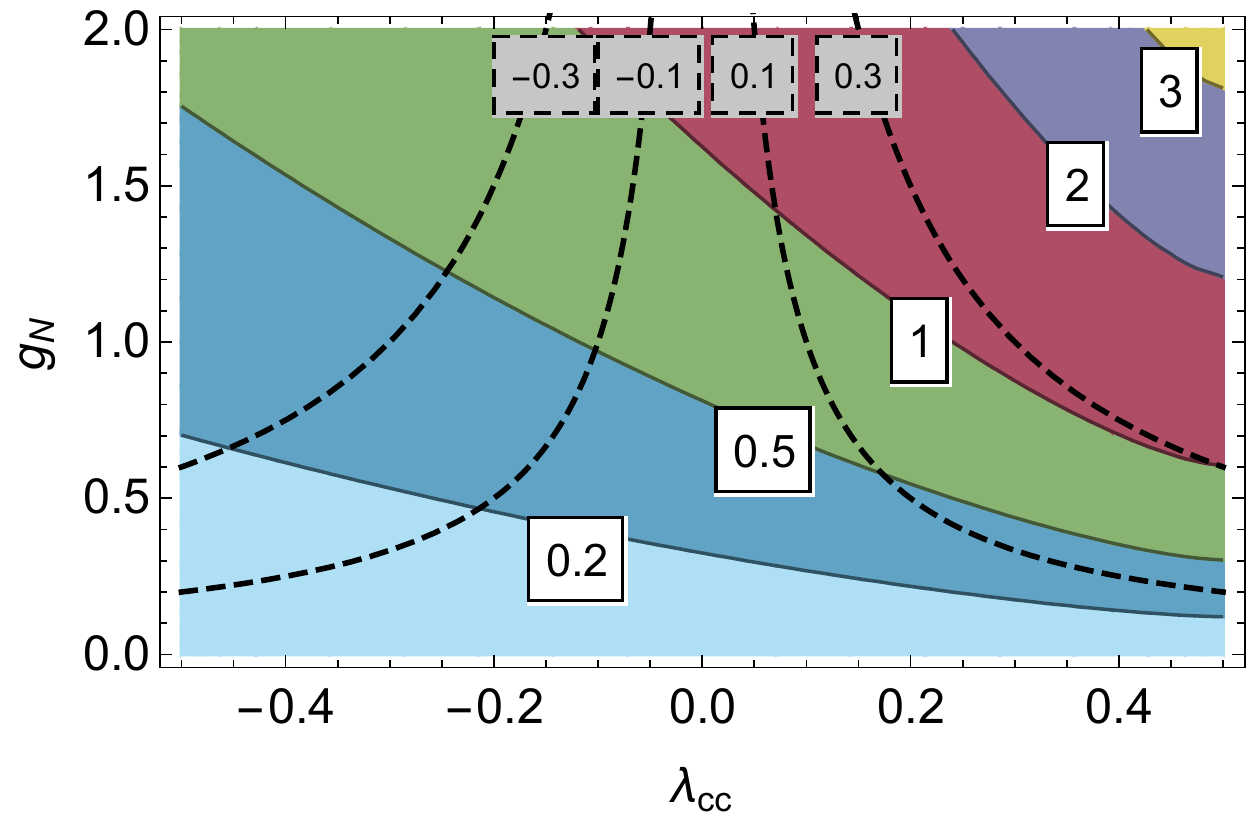}
\caption{\label{fig:eta4Fcontours} We show contours of $\eta_{4\rm F}$ in the $g_{\rm N}, \lambda_{\rm cc}$ plane. Dashed contours are different values of $\Lambda= g_{\rm N}\lambda_{\rm cc}$.}
\end{figure}

\emph{Dynamical suppression of proton decay in quantum gravity.}\\
A non-zero quantum-gravity contribution to the scaling dimension changes the relation between proton lifetime and $M_X$, because the beta function Eq.~\eqref{eq:betafunction} can, for (approximately) constant $\eta_{4 \rm F}$, be integrated\footnote{For non-constant $\eta_{4 \rm F}$, a numerical integration is possible; the case of constant $\eta_{4 \rm F}$ is instructive for the more general case.} to yield
\be
G_{4\rm F}^{qqq\ell}(M_X) = G_{4\rm F}^{qqq\ell}(k_{\rm UV}) \frac{M_X^{2+ \eta_{4\rm F}}}{k_{\rm UV}^{2+\eta_{4\rm F}}},
\ee
where $k_{\rm UV} > M_X$ is the UV scale, beyond which we assume that our quantum-field theoretic description breaks down.
In this way, because $\eta_{4 \rm F}>0$, we obtain
\be
G_{4\rm F}^{qqq\ell}(M_X)<1,
\ee
even if $G_{4\rm F}^{qqq\ell}(k_{\rm UV}) \approx 1$. How strong the suppression of $G_{4\rm F}^{qqq\ell}(M_X)$ is, depends on the values of $g_{\rm N}$ and $\lambda_{\rm cc}$. We illustrate the dependence of $\eta_{4\rm F}$ on $g_{\rm N}$ and $\lambda_{\rm cc}$ in Fig.~\ref{fig:eta4Fcontours}.

\begin{figure}
\includegraphics[width=\linewidth]{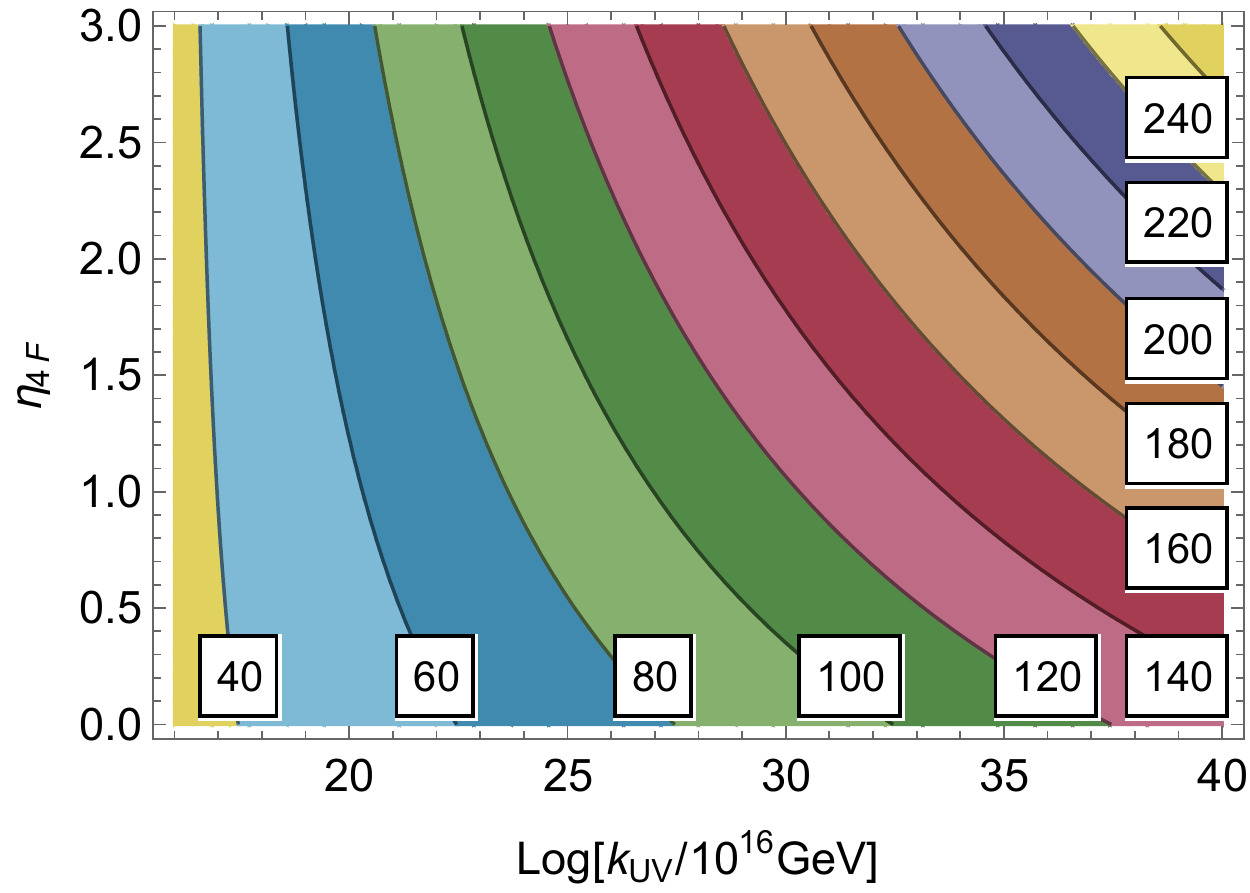}
\\
\includegraphics[width=\linewidth]{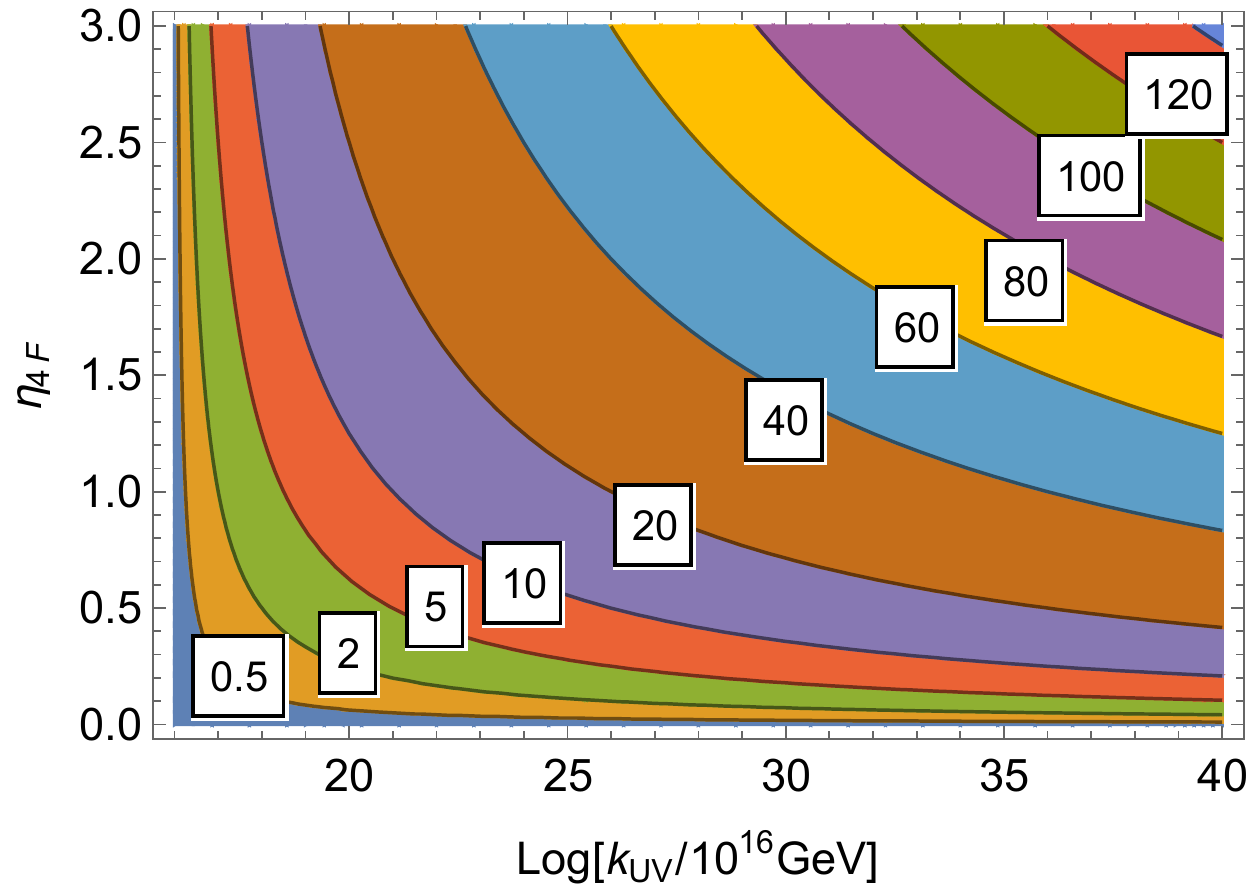}
\caption{\label{fig:taucontours} We show contours of ${\rm Log}_{10}(\tau_p)$ as a function of $k_{\rm UV}$ and $\eta_{4\rm F}$ (upper panel), and ${\rm Log}_{10}(\tau_p/\tau_{p,\, \rm canonical})$, where $\tau_{p,\, \rm canonical}$ is the proton lifetime in a setting with canonical scaling, i.e., with $\eta_{4 \rm F}=0$ (lower panel). For both panels, we set $M_{X}=10^{16}\, \rm GeV$ and $G_{4\rm F}^{qqq\ell}(k_{\rm UV})=1$.}
\end{figure}

Accordingly, we obtain a larger proton lifetime than expected based on the canonical scaling argument presented in Eq.~\ref{eq:protontauQGcanonical}, namely
\be
\tau_p \approx 16 \pi M_{p}^{-1} \left(G_{4\rm F}^{qqq\ell}(k_{\rm UV}) \frac{M_X^{2+ \eta_{4\rm F}}}{k_{\rm UV}^{2+\eta_{4\rm F}}} \right)^{-2} \frac{M_X^4}{M_p^4}.
\ee
We show contours of $\tau_p$ in the $(\eta_{4 \rm F},k_{\rm UV})$ plane in Fig.~\ref{fig:taucontours}. Even at modest values $k_{\rm UV} \approx 10^{18}\, \rm GeV$, proton lifetimes of order $10^{40}\, \rm years$ can be reached, putting the theory well on the safe side of the encroaching experimental bounds.

Within the quantum-gravity theory that holds above $k_{\rm UV}$, global symmetries are generically expected to be broken, with explicit studies confirming the expectation within string-theoretic and holographic settings \cite{Banks:1988yz,Giddings:1987cg,Kallosh:1995hi,Arkani-Hamed:2006emk,Banks:2010zn,Harlow:2018jwu,Harlow:2018tng}.\footnote{Other quantum-gravity settings may well be exceptions to this expectation, see, e.g., \cite{Eichhorn:2022gku}.} This includes baryon-number symmetry\footnote{Baryon number symmetry is actually anomalous in the Standard Model, but the anomaly does not give rise to a four-fermion interaction leading to proton decay, because it violates baryon number by 3 \cite{tHooft:1976rip}.  The contribution to proton decay is then through 12-fermion operators (= mass dimension 10); combined with instantonic prefactors as a function of the weak gauge coupling, the decay rate due to this effect is of order $10^{-173}$, see, e.g., \cite{Espinosa:1989qn} for details.} which, when violated, generically leads to proton decay. We thus expect that at $k_{\rm UV}$, $G_{4\rm F}^{qqq\ell}(k_{\rm UV})\approx 1$. For low enough $k_{\rm UV}$, this may even put quantum gravity theories in conflict with the experimental bounds, see \cite{Adams:2000za} for an example. We find that the situation is improved by accounting for quantum gravity fluctuations within the SMEFT regime, because $G_{4\rm F}^{qqq\ell}(M_X)\ll 1$ holds quite generically.

How strongly proton decay is suppressed, depends on the details of the quantum gravity setting, which determines $G_{4\rm F}^{qqq\ell}(k_{\rm UV})$ and $k_{\rm UV}$. Even for $\eta_{4\rm F} \approx 0.1$, i.e., a contribution well in the perturbative regime and the choice $k_{\rm UV} = 10\, M_{\rm Planck}$, $\tau_p$ is increased by nearly an order of magnitude.
The probably most conservative assumption puts $k_{\rm UV} = M_{\rm Planck}$ and takes the low-energy values for $G_{\rm N}$ and $\lambda_{\rm cc}$, i.e., neglects any running of the gravitational couplings. We thus set $\lambda_{\rm cc}=0$ and $g_{\rm N}= k^2/M_{\rm Planck}^2$ and obtain that $\eta_{4 \rm F} \sim k^2$, i.e., the gravitational contribution is only active over very few orders of magnitude below $M_{\rm Planck}$. Even so, we find an increase of $\tau_p$ by roughly a factor 2 by numerically integrating Eq.~\eqref{eq:betafunction}.
This shows that even for quantum gravity theories which set in as low as $M_{\rm Planck}$, the contribution  to the proton lifetime from the effective field theory regime is not negligible, such that any comparison to the experimental data should account for this contribution. It also means that quantum-gravity theories dynamically protect the stability of the proton -- at least in settings where four-fermion operators provide the relevant decay channel.

\paragraph{Implications for asymptotically safe quantum gravity:}
There is one particular quantum-gravity approach that serves as a special case of our general analysis, namely the asymptotic-safety scenario for quantum gravity and matter. In short, asymptotic safety relies on an enhanced symmetry in the UV, namely quantum scale symmetry. It is realized if all beta functions of the theory become zero. As a consequence, the quantum field theoretic description of gravity and matter can be extended to arbitrarily high energy scales.

Based on a proposal by Weinberg \cite{Weinberg:1980gg} and a methodology pioneered by Reuter \cite{Reuter:1996cp}, the modern asymptotic-safety approach has achieved a number of compelling results over the last decade: an asymptotically safe fixed point has been convincingly established in the Euclidean quantum gravity regime, see \cite{Bonanno:2020bil,Saueressig:2023irs} for reviews and references therein; and extended to include SM matter \cite{Eichhorn:2022gku}. Its predictive power has been established by showing how the ratio of Higgs mass to the electroweak scale and the ratio of the top mass to the electroweak scale can be calculated from first principles \cite{Shaposhnikov:2009pv,Eichhorn:2017ylw,Eichhorn:2018whv}, and the Abelian gauge coupling can be bounded from above \cite{Eichhorn:2017lry}. Moreover, indications for Lorentzian asymptotic safety and for the avoidance of unitarity-violation have been found \cite{Manrique:2011jc,Draper:2020bop,Fehre:2021eob}; evidence for singularity resolution has been discovered \cite{Lehners:2019ibe,Bosma:2019aiu,Borissova:2020knn} and an argument why the spacetime dimensionality is in fact 4 has been advanced \cite{Eichhorn:2019yzm}.

In this approach, the quantum field theoretic treatment can be extended to $k_{\rm UV} \rightarrow \infty$. The value of $G_{4\rm F}^{qqq\ell}$ is determined by quantum scale symmetry.  Thus, we first ask whether there is a (non-trivial) zero of the beta function $\beta_{G_{4\rm F}^{qqq\ell}}$. 
We find that there are no terms $\sim g_{\rm N}^{\alpha}\, (G_{4\rm F}^{qqq\ell})^{0}$, $\alpha=1,2,...$ in $\beta_{G_{4\rm F}^{qqq\ell}}$, see the supplementary material for details. Therefore, we conclude that there is a trivial zero of $\beta_{G_{4\rm F}^{qqq\ell}}$. This is an important difference to other four-fermion couplings, which have been analyzed in \cite{Eichhorn:2011pc}, see also \cite{Meibohm:2016mkp,Eichhorn:2017eht,deBrito:2020dta}, and for which gravity shifts the trivial zero to a non-trivial one. The difference can be traced back to the underlying symmetries: asymptotically safe gravity generates those four-fermion interactions in the SMEFT, which can roughly be written as a product of two kinetic terms. As a consequence, no four-fermion operator with three quarks and one lepton is generated. Thus, the main effect of gravity on $G_{4\rm F}^{qqq\ell}$ remains the anomalous scaling dimension $\eta_{4 \rm F}$. Because $2+\eta_{4 \rm F}>0$ holds in asymptotic safety\footnote{One may supplement our analysis by beta functions for the gravitational couplings, which, under the impact of Standard Model matter fields yield fixed-point values $g_{\rm N} = 3.3$ and $\lambda_{\rm cc} = -4.5$, see \cite{Eichhorn:2017ylw}, such that $\eta_{4 \rm F}=0.07$.}, the four-fermion coupling is irrelevant at the asymptotically safe fixed point. In other words, quantum gravity fluctuations drive the coupling back to zero, even if it is by hand set away from the fixed-point value.
As a consequence, the asymptotically safe prediction is
\be
G_{4\rm F}^{qqq\ell}(k=M_{\rm Planck})=0,
\ee
where $M_{\rm Planck}$ is the scale at which quantum-gravity fluctuations in asymptotic safety become negligible (roughly $M_{\rm Planck} = 10^{19}\, \rm GeV$). From this it follows that
\be
\tau_p \rightarrow \infty,
\ee
in the asymptotically safe Standard Model with gravity.\footnote{Instanton effects in the asymptotically safe path integral may become important, so that the proton may not be completely stable, see footnote 3.} {This result holds within our approximation of truncating the gravitational dynamics to the Einstein-Hilbert action, which is expected to capture the leading gravitational dynamics.}
In asymptotically safe grand unified theories, the situation is different, because a nonzero four-fermion coupling is present. An upper bound of $\left(G_{4\rm F}^{qqql} \right)^2\lesssim 10^{-7}$ for its value can be predicted in asymptotic safety, see the supplementary material.

Before concluding, we highlight a main assumption of our analysis: we  assume a near-perturbative gravitational regime, where strongly non-perturbative effects, encoded in higher-order operators, are negligible, such that we can parameterize metric fluctuations in terms of the Einstein-Hilbert action. There is support for such a near-perturbative regime from several directions, including an analysis of scaling dimensions \cite{Falls:2013bv,Falls:2014tra,Falls:2017lst,Falls:2018ylp}, non-trivial symmetry identities \cite{Eichhorn:2018akn,Eichhorn:2018ydy}, a weak-gravity bound on the size of gravitational effects \cite{Eichhorn:2016esv,Christiansen:2017gtg,deBrito:2021pyi,Eichhorn:2021qet,deBrito:2023myf} and explicit evaluations of the size of the quantum gravitational effects on matter, cf., e.g., \cite{Eichhorn:2017ylw,Eichhorn:2020sbo,Eichhorn:2022gku}. {In such a regime, \cite{Eichhorn:2017eht} has shown explicitly that the quantum-gravity contribution to the scaling dimension of couplings dominates over additional effects, such as, e.g., contributions from induced matter self-interactions.}
Nevertheless, it is not in principle excluded that effects neither included in the previous studies nor the present work change this.
However, our results are actually consistent with the assumption of near-perturbativity, because the fixed-point value of the gravity-induced correction to scaling dimension is in fact $\eta_{4\rm F\, \ast}\approx 0.07$.

\emph{Summary and outlook}:\\
We have shown that quantum gravity dynamically protects the stability of the proton by increasing the scaling dimension of four-fermion operators which mediate proton decay. Near-future experiments will provide a nontrivial test of our prediction, because the experimental bounds on the proton lifetime can be translated into constraints on the value of $\eta_{4\rm F}$. 

It is also interesting that quantum-gravity fluctuations in the effective-field-theory setting have the opposite effect of what is generically expected of quantum gravity: it is expected that quantum gravity breaks global symmetries, which would imply that proton-decay-mediating four-fermion operators are generated by quantum gravity. Our explicit calculation shows the opposite, namely that quantum gravity fluctuations dynamically suppress these operators.

We finally highlight that if experiments discover the proton to be unstable, the asymptotically safe Standard Model with gravity comes under pressure: with the caveat that large nonperturbative effects would invalidate our result, we predict a very large proton lifetime well out of reach of near-future experiments.

\begin{acknowledgments}
We thank A.~Held and G.~P.~de Brito for helpful comments and discussions. %
This research is supported by a research grant (29405) from VILLUM Fonden and by the Deutsche Forschungsgemeinschaft (DFG) through the Walter Benjamin program (RA3854/1-1, Project id No. 518075237), the SFB 1143 (Project A07, Project id No. 247310070), the Würzburg-Dresden Cluster of Excellence ct.qmat (EXC2147, Project id No. 390858490) and the Emmy Noether program (JA2306/4-1, Project id No. 411750675).
\end{acknowledgments}

%

\end{document}


\title{
Supplemental Material:\\
Suppression of proton decay in quantum gravity}
 
 \author{Astrid Eichhorn}
   \email{eichhorn@cp3.sdu.dk}
\affiliation{CP3-Origins, University of Southern Denmark, Campusvej 55, 5230 Odense M, Denmark}
\author{Shouryya Ray}
\email{sray@cp3.sdu.dk}
\affiliation{CP3-Origins, University of Southern Denmark, Campusvej 55, 5230 Odense M, Denmark}
\affiliation{Institut f\"ur Theoretische Physik and W\"urzburg-Dresden Cluster of Excellence ct.qmat, TU Dresden, 01062 Dresden, Germany}


\date{\today}

\maketitle


\section{Model and method}
\label{sec:framework:model}
The aim of this section is to define the framework within which the computations in the main text were performed.

To begin with, let us define our model. The matter sector consists of one Dirac spinor $\psi$, encompassing all fermionic matter degrees of freedom in the Standard Model plus three right-handed neutrinos. The fermions in the SM are Weyl fermions, some of which transform in the fundamental representation of the $\operatorname{SU}(3)_\text{colour}$ and $\operatorname{SU}(2)_\text{L}$ gauge groups. Gravity is `blind' with respect to such internal indices. We can therefore use a compact notation in which we group all SM fermions into one spinor $\psi_{\alpha i}$ that is a collection of Dirac spinors formed by combining the left- and right-handed Weyl fermions of the SM (ameliorated by right-handed neutrinos). %
The Dirac index is denoted $\alpha$ (and suppressed wherever possible), whilst $i$ is a collection of all other internal indices. We do not require the index $i$ of $\psi$ to transform in a simple way under the gauge groups of the SM, because such transformation properties are immaterial for the interplay with gravity. %

The index $i$ may be understood as a combined generation, SU(3) colour, and SU(2) isospin index.
In the SM, the symmetry between left- and right-handed fermions is broken by a nonvanishing $\operatorname{SU}(2)_\text{L}$ gauge coupling. Within the asymptotically safe SM with gravity, asymptotic freedom of the $\operatorname{SU}(2)_\text{L}$ gauge coupling persists  \cite{Daum:2009dn,Folkerts:2011jz,Christiansen:2017cxa} and thus, at high energies, left-right symmetry is restored asymptotically. We neglect the small breaking of this symmetry that occurs even at transplanckian scales, when the $\operatorname{SU}(2)_\text{L}$ gauge coupling increases from zero towards its nonzero value at the Planck scale.

The equal number of left- and right-handed fermions also ensures that the gravitational contribution to the axial anomaly does not percolate into a lepton number anomaly. The asymptotic left-right symmetry that emerges in the limit of vanishing $\operatorname{SU}(2)_\text{L}$ gauge coupling allows us to form \emph{Dirac} $\operatorname{SU}(2)$-doublets, i.e., $q = \left(\begin{smallmatrix}d \\ u\end{smallmatrix}\right)$ and $\ell = \left(\begin{smallmatrix}e \\ \nu\end{smallmatrix}\right)$. It is expedient to group the three quark colours $q = (q_\text{r}, q_\text{g}, q_\text{b})$ along with the lepton $\ell$ to form
\begin{align}
    (\psi_i) = \begin{pmatrix}
        q_\text{r} \\
        q_\text{g} \\
        q_\text{b} \\
        \ell
    \end{pmatrix}.
\end{align}
In this arrangement, the index $i$ with $i \in \{1,2,3,4\}$ is often referred to as the Pati--Salam colour index. Processes involving quark-lepton transmutation are encoded in non-vanishing matrix elements between the $i \leqslant 3$ and $i=4$ subspaces, which makes the Pati--Salam index a natural language for discussing proton decay.

The Euclidean action for fermions minimally coupled to gravity reads
\begin{align}
    S_{\text{kin},\text{F}} = i \int_x \sqrt{g} \bar{\psi}_i \slashed{\nabla} \psi_i.
\end{align}
The coupling to the spacetime metric arises through the determinant factor $\sqrt{g}$ where $g = \operatorname{det}(g_{\mu\nu})$, as well as the covariant derivative $\nabla_\mu$, which contains the spin connection in the case of spin-1/2 fields such as $\psi$. The dynamics of the metric are described by the Einstein--Hilbert action
\begin{align}
    S_\text{EH} = \frac{1}{16\pi G_\text{N}} \int_x \sqrt{g} \left(2\Lambda_\text{cc} - R\right),
\end{align}
where $R$ is the Ricci scalar, $\Lambda_\text{cc}$ is the cosmological constant and $G_{\rm N}$ is the Newton coupling.

\begin{figure*}
    \includegraphics[width=\textwidth]{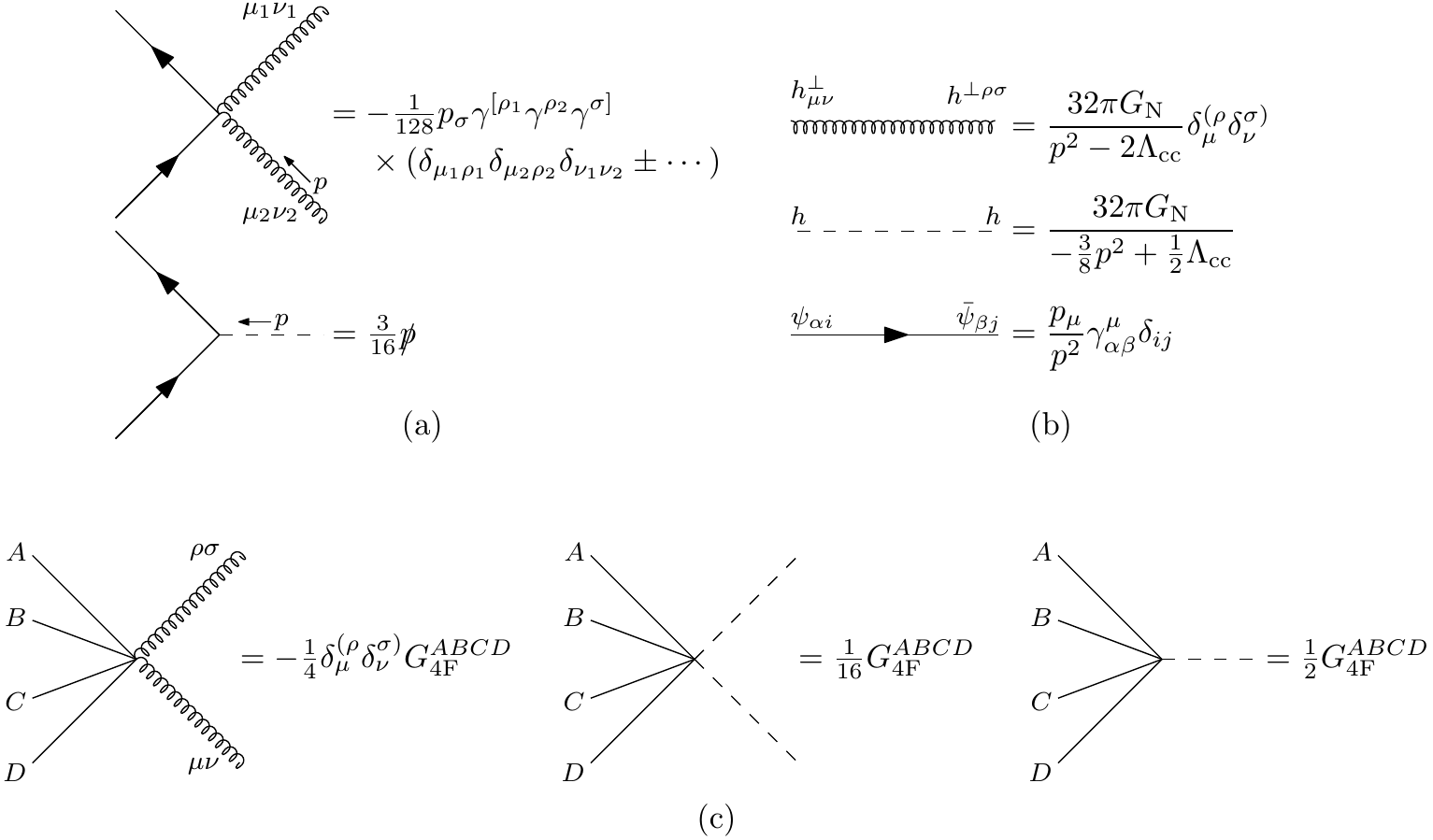}
    \caption{Propagators and vertices arising from the classical action $S = S_\text{EH} + S_{\text{kin},\text{F}} + S_{4\text{F}}$: (a) minimal coupling of free fermions to metric fluctuations, (b) free fermion and metric fluctuation propagators, (c) vertices coupling fermions and metric fluctuations due to the determinant factor $\sqrt{g}$ in $S_{4\text{F}}$. Here, the metric fluctuation field $h_{\mu\nu}$ is defined as a linear split of the metric $g_{\mu\nu}(x) = \bar{g}_{\mu\nu}(x) + h_{\mu\nu}(x)$ with background metric $\bar{g}_{\mu\nu}$; we have only kept up to quadratic terms in $h_{\mu\nu}$ and set $\bar{g}_{\mu\nu} \to \delta_{\mu\nu}$ when noting vertices and propagators. We are working in the Landau--DeWitt gauge, so that only transverse traceless $h_{\mu\nu}^\perp$ and trace $h$ modes can contribute. Solid lines without an arrow denote Nambu--Gor'kov spinors, see Eq.~\eqref{eq:NambuGorkov}. Due to the Grassmann nature of $\Psi$, the $G_{4\text{F}}^{ABCD}$ may be assumed to be completely antisymmetric in $ABCD$ w.l.o.g.}
    \label{fig:vertices}
\end{figure*}

We work within the background field formalism, such that an auxiliary background metric is available to gauge fix and also to set up a Renormalization Group (RG) flow \cite{Reuter:1996cp,Reuter:2019byg}. %
We thus expand the metric as $g_{\mu\nu} = \bar{g}_{\mu\nu} + h_{\mu\nu}$, where $h_{\mu\nu}$ are the fluctuations around the auxiliary background metric $\bar{g}_{\mu\nu}$. We are not working in a perturbative framework; thus there is no restriction in the amplitude of $h_{\mu\nu}$. The path integral over metric fluctuations $g_{\mu\nu}$ can thus be rewritten in terms of the path integral over $h_{\mu\nu}$. %
This leads to the vertices shown in Fig.~\ref{fig:vertices}(a), where we have set $\bar{g}_{\mu\nu} \to \delta_{\mu\nu}$ for simplicity (it is sufficient for our purposes, since we are ultimately interested in the RG flow of 4-Fermi couplings). Propagators have the standard form, summarized in Fig.~\ref{fig:vertices}(b); since we work in the Landau--DeWitt gauge, only the transverse traceless mode $h_{\mu\nu}^\perp$ and the trace mode $h$ contribute.

Finally, to describe proton decay, we need dimension 6 operators of the 4-Fermi type,
\begin{align}
    S_\text{4F} = \frac{1}{4!}G_{4\text{F}}^{ABCD}\int_x \sqrt{g} \Psi_A \Psi_B \Psi_C \Psi_D. \label{eq:S4F}
\end{align}
Here, 
\begin{align}
    \Psi \equiv (\Psi_A) = \begin{pmatrix} 
    \psi \\ \psi^c
    \end{pmatrix}
    \label{eq:NambuGorkov}
\end{align}
is the Nambu--Gor'kov spinor constructed from $\psi$, with $\psi^c$ the charge conjugate of $\psi$; the $A$'s are to be understood as a combined index collecting Nambu--Gor'kov isospin along with the usual Dirac spinor and other SM indices. As pointed out in \cite{Eichhorn:2011pc}, only the determinant factor $\sqrt{g}$ can lead to fermion-graviton vertices; these are summarized in Fig.~\ref{fig:vertices}(c). The term \eqref{eq:S4F} clearly describes every 4-Fermi operator conceivable; the specific forms of the coefficients $G_{4\text{F}}^{ABCD}$ that describe proton decay operators have been tabulated in the SMEFT literature, cf., e.g., \cite{Grzadkowski:2010es} for an exhaustive list. We shall denote all such couplings schematically (in a slight abuse of notation) as $G_{4\text{F}}^{qqql}$; it will in fact turn out that the metric fluctuation contributions to the scaling dimension of these operators are independent of the specific structure of the coefficients $G_{4\text{F}}^{ABCD}$.

For completeness, let us mention that we shall be deriving the flow equations using the functional renormalization group (FRG) approach, see \cite{Dupuis:2020fhh}. The flow equation is \cite{Wetterich:1992yh,Ellwanger:1993mw,Morris:1993qb}
\begin{align}
    k \partial_k \Gamma_k = \frac12 \operatorname{STr}\frac{k\partial_k R_k}{\Gamma_k^{(2)} + R_k},
\end{align}
where the IR cut-off is implemented by a bilinear term with kernel $R_k$. The supertrace $\operatorname{STr}$ includes an integration over coordinates as well as summation over indices; the `super' refers to the minus sign when tracing over the fermionic sector.

A useful way to re-write the FRG flow equation is in the form
\begin{align}
    k \partial_k \Gamma_k = \frac{1}{2} \operatorname{STr} k\tilde{\partial}_k \ln \mathcal{P}_k + \frac{1}{2} \sum_{n=0}^{\infty} \frac{(-1)^{n-1}}{n} \operatorname{STr}k\tilde{\partial}_k\left(\mathcal{P}_k^{-1} \mathcal{F}_k\right)^n.
    \label{eq:FRGmaster}
\end{align}
Here, $\tilde{\partial}_k$ only acts on the regulator dependence,
\begin{align}
    \tilde{\partial}_k = \int \partial_k R_k \frac{\delta}{\delta R_k},
\end{align}
and $\Gamma_k^{(2)} + R_k$ has been split into the regularized inverse propagator $\mathcal{P}_k$ and the field-dependent part $\mathcal{F}_k$ (roughly speaking the vertices of the theory) as
\begin{align}
    \mathcal{P}_k(-i\partial_\mu)\,\delta(x - y) &= \left.\Gamma_k^{(2)}\right|_{h_{\mu\nu} = \Psi_A = 0}(x,y) + R_k(x,y), \\
    \mathcal{F}_k(-i\partial_\mu)\,\delta(x - y) &= \Gamma_k^{(2)}(x,y) - \left.\Gamma_k^{(2)}\right|_{h_{\mu\nu} = \Psi_A = 0}(x,y).
\end{align}
Eq.~\eqref{eq:FRGmaster} makes the 1-loop nature manifest. In fact, up to the $\tilde{\partial}_k$, the contributions on the right-hand side are simply those given by 1PI one-loop diagrams, but with vertices and propagators replaced by their dressed avatars. The effect of the $\tilde{\partial}_k$ is to replace the loop integral by so-called threshold functions, where a diagram with $n_\text{F}$ fermion lines, $n_\perp$ TT lines, $n_\text{t}$ trace lines is associated with the threshold function $I_{n_\text{F},n_\perp,n_\text{t}}$. We shall be using the Litim regulator, for which the threshold functions are known analytically \cite{Eichhorn:2011pc}.

The essence of the approximation is then contained in the choice of ansatz for $\Gamma_k$. We shall be working in the minimal truncation, defined by the ansatz \footnote{If one were to compute physical observables by taking functional derivatives of the $\Gamma_k$ expressed in Eq.~\eqref{eq:mintruncation}, explicit appearances of the wavefunction renormalization $Z_{\text{N},\text{F}}$ would drop out (essentially because the coefficient of the kinetic operator has to be normalized to unity). Such couplings are called \emph{inessential} couplings and do not need to obey fixed-point equations of their own. Physical observables in general have to be invariant under field re-definitions, though making this invariance manifest in the ansatz can be non-trivial for general (non-linear) field re-definitions. A systematic procedure for eliminating dependencies on general field re-definitions has been formulated recently by Baldazzi \& Falls \cite{Baldazzi:2021ydj}, see also \cite{Baldazzi:2021orb,Knorr:2022ilz} for applications to quantum gravity. For our purposes, the rescaling of fields by constants (= wavefunction renormalization) is sufficient, and can be treated at the level of the ansatz, as we do hereinafter.} for the running effective action
\begin{align}
    \Gamma_k[h_{\mu\nu},\Psi_A] = S\!\left[\sqrt{Z_\text{N}}(k) h_{\mu\nu}, \sqrt{Z_\text{F}}(k) \Psi_A ; G_\text{N}(k), \Lambda_\text{cc}(k)\right],
    \label{eq:mintruncation}
\end{align}
which means the vertices and propagators entering Eq.~\eqref{eq:FRGmaster} are precisely the dressed (and IR-regulated if applicable) versions of those appearing in Fig.~\ref{fig:vertices}. The definitions of the dimensionless couplings are
\begin{align}
    g_\text{N}(k) &= k^{2} G_\text{N}(k), \\
    \lambda_\text{cc}(k) &= k^{-2}\Lambda_\text{cc}(k), \\
    g_\text{4F}^{ABCD}(k) &= k^{2} G_\text{4F}^{ABCD}(k).
\end{align}

\section{Ward-Takahashi identity for $B$-symmetry}
\begin{figure*}
    \centering
    \includegraphics[scale=1]{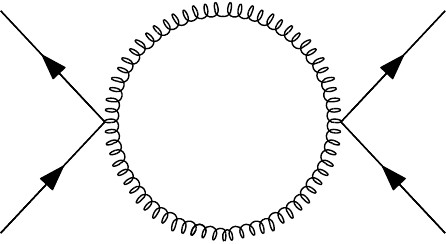}
    \caption{`Candy' diagram -- the only diagram with four external fermion legs and no 4-Fermi vertices. The curly line indicates metric fluctuations.}
    \label{fig:candy}
\end{figure*}

\begin{widetext}
Here we consider whether $B$-violating interactions are generated by metric fluctuations. At the level of 4-Fermi operators, the only diagram at $\mathcal{O}((g_\text{4F})^0)$ is the `candy' diagram shown in Fig.~\ref{fig:candy},
\begin{align}
    \left[\text{Fig.~\ref{fig:candy}}\right] = -\frac{1}{2} \mathcal{F}_{\text{kin}\:k}^{h_{\mu_1 \mu_2}^\perp h_{\mu_3 \mu_4}^\perp} \left(\mathcal{P}_k^{-1}\right)_{h_{\mu_3 \mu_4}^\perp h_{\mu_5 \mu_6}^\perp} \mathcal{F}_{\text{kin}\:k}^{h_{\mu_5 \mu_6}^\perp h_{\mu_7 \mu_8}^\perp} \left(\mathcal{P}_k^{-1}\right)_{h_{\mu_7 \mu_8}^\perp h_{\mu_1 \mu_2}^\perp},
\end{align}
where $\mathcal{F}_\text{kin}$ refers to the vertices arising from the expansion of the $g_{\mu\nu}$-dependence of $S_\text{kin}$ in powers of $h_{\mu\nu}$. Acting with $k\tilde{\partial}_k$ and expressed using the threshold functions of \cite{Eichhorn:2011pc}, one finds
\begin{align}
    k\tilde{\partial}_k\left[\text{Fig.~\ref{fig:candy}}\right] = \frac{15}{256\times 16}I_{020} \left(\bar\psi_a \gamma^\mu {  \gamma^5} \psi_a\right)^2,
\end{align}
in agreement with \emph{ibid.} %
\end{widetext}
Each of the two bilinears that constitute the four-fermion operator, is diagonal in the Pati--Salam index. There are therefore four-fermion operators that mix quarks with leptons, but they contain an even number of each and therefore do not result in proton decay.  %
This may be summarized by saying
\begin{align}
    k\partial_k g_{4\text{F}}^{qqql} = \mathcal{O}(g_{4\text{F}}^{qqql}).
    \label{eq:noprotondecay}
\end{align}
In SMEFT (i.e., without gravity), the validity of this relation has been established previously \cite{Abbott:1980zj,Alonso:2014zka}. In other words, within the SM, $B$-violating 4-Fermi operators are not generated by quantum fluctuations. Our calculation shows that this is respected by gravity fluctuations.

\begin{widetext}
To go beyond 4-Fermi operators, consider now the Euclidean path integral for $\Gamma_k$ \cite{Baldazzi:2021ydj,Baldazzi:2021orb},
\begin{align}\label{eq:Gammakdef}
    e^{-\Gamma_k[\Phi]} = \int \mathcal{D}\tilde{\Phi}\,e^{-S[\tilde{\Phi}] + (\tilde{\Phi}_X - \Phi_X) \Gamma_{k,}^{\hphantom{k,}X}[\Phi] - \frac12 \mathcal{R}_k^{XY}(\tilde{\Phi}_X - \Phi_X) (\tilde{\Phi}_Y - \Phi_Y)},
\end{align}
where $X$ is a DeWitt index for $(\Phi_X) = \left(\begin{smallmatrix} h_{\mu\nu}(x) \\ \Psi_A(x) \end{smallmatrix}\right)$. We use $\tilde{\Phi}_X$ to denote the integration variable, to be distinguished from the expectation value $\Phi_X$.
Consider now an infinitesimal $\operatorname{U}(1)_\text{B}$ transformation
\begin{align}
\delta_{\epsilon_\text{B}} \Phi_X = i \epsilon_\text{B} Q_{\text{B}}^{XY} \Phi_Y
\end{align}
where $Q_{\text{B}}^{XY}$ is diagonal and equals $1/3$ ($-1/3$) for the (anti-)quark entries and zero otherwise. The variation of $\Gamma_k[\Phi]$ is
\begin{align}
    \delta_{\epsilon_\text{B}} \Gamma_k[\Phi] = \left\langle \delta_{\epsilon_\text{B}} \left(S[\tilde\Phi] - (\tilde{\Phi}_X - \Phi_X) \Gamma_{k,}^{\hphantom{k,}X}[\Phi] + \frac12 \mathcal{R}_k^{XY}(\tilde{\Phi}_X - \Phi_X) (\tilde{\Phi}_Y - \Phi_Y)\right) \right\rangle_{k;\Phi}
\end{align}
with $\left\langle \mathcal{F}[\tilde{\Phi}] \right\rangle_{k;\Phi} = e^{\Gamma_k[\Phi]}\int \mathcal{D}\tilde{\Phi} e^{-S[\Phi] + (\tilde{\Phi}_X - \Phi_X)\Gamma_{k,}^{\hphantom{k,}X}[\Phi] - \frac12 \mathcal{R}_k^{XY}(\tilde{\Phi}_X - \Phi_X) (\tilde{\Phi}_Y - \Phi_Y)} \mathcal{F}[\tilde{\Phi}]$; note that this implies $\langle \tilde{\Phi} \rangle_{k;\Phi} = \Phi$. The variations of both the classical action and the regulator term manifestly vanish under $\delta_{\epsilon_\text{B}}$. Furthermore, since $\langle \tilde{\Phi}_X - \Phi_X \rangle_{k;\Phi} = 0$ and $\delta_{\epsilon_\text{B}}$ is a derivation (i.e., is linear and obeys the product rule), the $\Gamma_{k,}^{\hphantom{k,}X}[\Phi]$-term on the right-hand side also vanishes upon taking the expectation value. 
Ultimately, this implies
\begin{align}
    \delta_{\epsilon_\text{B}} \Gamma_k[\Phi] = 0,
    \label{eq:noprotondecaygen}
\end{align}
which is the truncation-independent generalization of Eq.~\eqref{eq:noprotondecay} as advertised. Let us note in passing that the derivation above would go through for any symmetry of $S_{\text{kin},\text{F}}$ that is realized linearly and not explicitly broken by the regulator. In this derivation, we assume that a UV-regularized path-integral measure can be written to make Eq.~\eqref{eq:Gammakdef} well-defined without violating the symmetry. See also \cite{Gies:2006wv,Laporte:2021kyp} for similar derivations.
\end{widetext}

\section{Computation of $\eta_{4\text{F}}$}
\begin{figure*}
    \centering
    \includegraphics[scale=1]{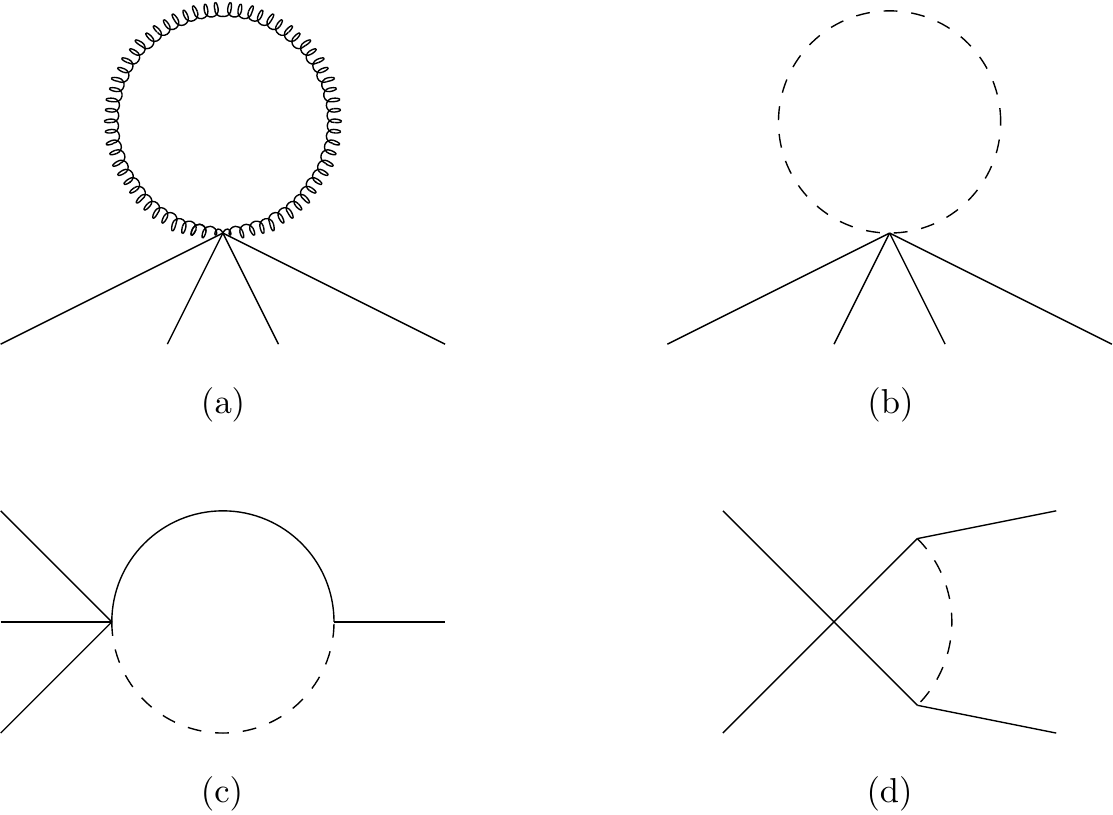}
    \caption{Metric fluctuation contributions to $\eta_{4\text{F}}$. Arrowless solid lines denote Nambu--Gor'kov spinors, see the discussion in Sec.~\ref{sec:framework:model}. Curly lines denote transverse traceless metric fluctuations and dashed lines denote the trace fluctuations.}
    \label{fig:eta4Fdiags}
\end{figure*}
\begin{widetext}
The diagrams contributing to $\eta_{4\text{F}}$ are shown in Fig.~\ref{fig:eta4Fdiags}. %
The results of diagrams (a) and (b) are manifestly proportional to $G_{4\text{F}}^{ABCD}$ times a factor independent of $G_{4\text{F}}^{ABCD}$. The more involved contributions are displayed in diagrams (c) and (d). Explicitly, the contractions can be written as
\begin{align}
\text{Fig.~\ref{fig:eta4Fdiags}(c)} &= -\frac{1}{4} \left[\vphantom{\mathcal{F}_\text{kin}^{\Psi_A \Psi_B}} \mathcal{F}_\text{kin}^{h\Psi_A} \left(\mathcal{P}^{-1}\right)_{\Psi_A \Psi_B} \mathcal{F}_{4\text{F}}^{\Psi_B h} \left(\mathcal{P}^{-1}\right)_{hh} \right. \nonumber\\
    & \hphantom{{}={}\big[}
    \left. 
    {}+ \mathcal{F}_\text{4F}^{h\Psi_{A}} \left(\mathcal{P}^{-1}\right)_{\Psi_{A}\Psi_{B}}  \mathcal{F}_\text{kin}^{\Psi_{B} h} \left(\mathcal{P}^{-1}\right)_{hh} \right. \nonumber\\
    & \hphantom{{}={}\big[}
    \left. 
    {}+ \mathcal{F}_{4\text{F}}^{\Psi_{A} h} \left(\mathcal{P}^{-1}\right)_{hh} \mathcal{F}_\text{kin}^{h \Psi_{B}} \left(\mathcal{P}^{-1}\right)_{\Psi_B \Psi_A} (-1) \right. \nonumber\\
    & \hphantom{{}={}\big[}
    \left. 
    {}+ \mathcal{F}_\text{kin}^{\Psi_A h} \left(\mathcal{P}^{-1}\right)_{hh} \mathcal{F}_\text{4F}^{h {\Psi}_{B}} \left(\mathcal{P}^{-1}\right)_{\Psi_B \Psi_A} (-1) \right], \label{eq:eta4Fcancel1} \\
    \text{Fig.~\ref{fig:eta4Fdiags}(d)} &= \frac{1}{6} \left[
    - \left(\mathcal{P}^{-1}\right)_{\Psi_A \Psi_{B}} \mathcal{F}_\text{kin}^{\Psi_B h} \left(\mathcal{P}^{-1}\right)_{hh} \mathcal{F}_\text{kin}^{h \Psi_{C}} \left(\mathcal{P}^{-1}\right)_{\Psi_{C} \Psi_{D}} \mathcal{F}_{4\text{F}}^{\Psi_D \Psi_{A}} \right. \nonumber\\
    & \hphantom{{}={}\big[}
    \left. 
    {}- \left(\mathcal{P}^{-1}\right)_{\Psi_{A} {\Psi}_{B}} \mathcal{F}_{4\text{F}}^{{\Psi}_{B} \Psi_{C}} \left(\mathcal{P}^{-1}\right)_{\Psi_{C} \Psi_{D}} \mathcal{F}_\text{kin}^{\Psi_{D} h} \left(\mathcal{P}^{-1}\right)_{hh} \mathcal{F}_{\text{kin}}^{h \Psi_{A}} \right. \nonumber\\
    & \hphantom{{}={}\big[}
    \left. 
    {}+ \left(\mathcal{P}^{-1}\right)_{hh} \mathcal{F}_\text{kin}^{h \Psi_{A}} \left(\mathcal{P}^{-1}\right)_{\Psi_{A} \Psi_{B}} \mathcal{F}_{4\text{F}}^{\Psi_{B} \Psi_{C}} \left(\mathcal{P}^{-1}\right)_{{\Psi}_{C} \Psi_{D}} \mathcal{F}_\text{kin}^{\Psi_{D} h} \right].
\end{align}
Using the antisymmetry relations for $G_{4\text{F}}^{ABCD}$, we can collapse the individual contributions to
\begin{align}
    \text{Fig.~\ref{fig:eta4Fdiags}(c)} &= -\left(\mathcal{P}^{-1}\right)_{hh} \mathcal{F}_\text{kin}^{h \Psi_A} \left(\mathcal{P}^{-1}\right)_{\Psi_{A}\Psi_{A'}} \mathcal{F}_\text{4F}^{{\Psi}_{A'}h}, \\
    \text{Fig.~\ref{fig:eta4Fdiags}(d)} &= \frac{1}{2} \left(\mathcal{P}^{-1}\right)_{hh} \left(\mathcal{P}^{-1}\right)_{\Psi_{A'} \Psi_{A}} \left(\mathcal{P}^{-1}\right)_{\Psi_{B'} {\Psi}_{B}} \mathcal{F}_\text{kin}^{\Psi_{A} h} \mathcal{F}_\text{kin}^{\Psi_{B} h} \mathcal{F}_{4\text{F}}^{\Psi_{A'} \Psi_{B'}}. \label{eq:eta4Fcancel-1}
\end{align}
Now, contracting a fermion propagator with a kinetic vertex always leads to a trivial structure in spinor space. Thus, the final result for $\eta_{4\text{F}}$ is independent \footnote{This independence of the index structure is also why in \cite{Eichhorn:2011pc} and follow-up works, the chirally symmetric four-fermion interaction $\left(\bar{\psi}\gamma_{\mu}\psi\right)^2$, although compatible with all symmetries, is not induced. The technically non-trivial cancellations needed to ensure this result are displayed in Eqs.~\eqref{eq:eta4Fcancel1}--\eqref{eq:eta4Fcancel-1}. In hindsight, the fact that every fermion propagator is `paired up' with a kinetic vertex may be considered intuitive, given that one usually expects gravity to be `blind with respect to internal indices'.}  of the specific index structure of $G_{4\text{F}}^{ABCD}$,
and works out to be
\begin{align}
    \eta_{4\text{F}} &= 2g_\text{N} \left[ \frac{5(6 - \eta_\text{N})}{24\pi (1 - \lambda_{\text{cc}})^2} + \frac{6 - \eta_\text{N}}{4\pi (- 3 + 4\lambda_{\text{cc}})^2} + \frac{36\eta_\text{N} - 7(54 - 24\lambda_\text{cc} + \eta_\text{F}(-3 + 4 \lambda_\text{cc}))}{35\pi(-3 + 4\lambda_\text{cc})^2}\right. \nonumber\\
    &\hphantom{{}={}} \left.{}- \frac{9(21\eta_\text{N} + 24(-14 + \eta_\text{F}) - 32 (-7 + \eta_\text{F})\lambda_\text{cc})}{448\pi (- 3 + 4\lambda_\text{cc})^2}
    \right],
\end{align}
in agreement with \cite{Eichhorn:2011pc}, where $\eta_\text{N} = -k\partial_k Z_\text{N}(k)$.
\end{widetext}

 In the following, we further set $\eta_\text{N} = 0$ and expand in powers of $\lambda_\text{cc}$, as done in the main text. This leads to the following result in terms of $\Lambda_{\rm cc}$:
\begin{align}
    \eta_{4\text{F}} = 2 G_\text{N} \sum_{n,l} c_{nl} \Lambda_\text{cc}^n \int \frac{\rmd^4 p}{(2\pi)^4} \frac{(p^2/k^2) r'(p^2/k^2)}{(p^2)^{m_n} [1 + r(p^2/k^2)]^{l}},
\end{align}
where $r$ is the dimensionless shape function associated with the regulator $R_k$, i.e., $R_k = p^2 r(p^2/k^2)$. By definition, $\eta_{4\text{F}}$ is a dimensionless quantity. The power $m_n$ is then fixed by dimensional analysis: $4 + 2m_n + 2n - 2 = 0$, whence $m_n = 1 + n$. The contribution proportional to the dimensionless combination $G_\text{N} \Lambda_\text{cc}$ corresponds to $n = 1$. In the main text, we asserted that it is regulator-independent. To see that it is indeed so, make the substitution $u \coloneqq p^2/k^2$; the integrals we are left with after integrating over the 3-sphere and modulo $r$-independent prefactors, are given by
\begin{align}
 I_{l} = \int_0^\infty \rmd u \frac{r'(u)}{[1 + r(u)]^l}.
\end{align}
The integrand can then be written as a total derivative, whence the regulator-independence follows from the universal UV- and IR-behavior of the shape function. In fact, $I_l=-\frac{1}{l-1}$, see \cite{Berges:2000ew,Codello:2008vh,deBrito:2022vbr}.


\section{Implications for asymptotically safe GUTs.}

 In typical GUT settings, quarks and leptons are grouped into one representation of the gauge group, such that the gauge bosons beyond the SM mediate proton decay. The resulting effective field theory contains proton-decay-mediating four-fermion operators with $M_X \approx M_{\rm GUT}$ corresponding to the mass of the gauge bosons beyond the SM. Additionally, proton decay can be mediated through a tree-level gauge boson exchange as well as dimension-five operators; depending on the GUT at hand. We focus on the four-fermion contribution, for which we show that it fulfils an upper bound, if the GUT is to be asymptotically safe.
 Schematically, the contribution to $\beta_{G_{4\rm F}^{qqq\ell}}$ from the gauge sector is given by
\be
\beta_{G_{4\rm F}^{qqq\ell}}\vert_{\rm GUT} =\#_{\rm GUT,\, 1}\,\frac{g_{\rm GUT}^4}{16\pi^2}+ \left(2+ \eta_{4\rm F} \right)G_{4\rm F}^{qqq\ell} + \mathcal{O}(G_{4\rm F}^{qqq\ell})^2,\label{eq:betafunctionGUT}
\ee
where $g_{\rm GUT}$ is the gauge coupling and $\#_{\rm GUT,\, 1}$ a numerical factor that depends on the gauge group. The scale dependence of the gauge coupling is also impacted by quantum gravity fluctuations \cite{Eichhorn:2017muy}, such that
\be
\beta_{g_{\rm GUT}} = - f_g\, g_{\rm GUT} + \frac{\#_{\rm GUT,\, 2}}{16\pi^2} g_{\rm GUT}^3,
\ee
with $f_g \geq 0$ the quantum-gravity contribution \cite{Daum:2009dn,Daum:2010bc,Folkerts:2011jz,Christiansen:2017gtg,Christiansen:2017cxa,DeBrito:2019gdd,Eichhorn:2021qet,deBrito:2022vbr} and $\#_{\rm GUT,\, 2}$ a factor that depends on the gauge group and the representations of fermion and scalar fields; we assume $\#_{\rm GUT,\, 2}>0$. From the beta function for the gauge coupling, one can infer that the value of the GUT-gauge coupling at the Planck scale, $g_{\rm GUT}(k=M_{\rm Planck})$ is bounded from above in asymptotic safety: The screening contributions of the GUT matter and gauge fields, encoded in $\#_{\rm GUT,\, 2}$, compete with the antiscreening contributions of gravity, encoded in $f_g$. For $g_{\rm GUT}=g_{\rm GUT, \, \ast} = \sqrt{16 \pi^2 f_g/\#_{\rm GUT,\, 2}}$, these balance out and generate an asymptotically safe fixed point. Values $g_{\rm GUT}(k=M_{\rm Planck})<g_{\rm GUT, \, \ast}$ are reachable from an asymptotically free fixed point. However, values $g_{\rm GUT}(k=M_{\rm Planck})>g_{\rm GUT, \, \ast}$ are not connected to any fixed point and result in Landau poles in the UV.

In turn, the asymptotically safe fixed point for the gauge coupling induces an asymptotically safe fixed point for the four-fermion coupling, 
\be
G_{4\rm F,\,\ast}^{qqq\ell}=-\frac{\#_{\rm GUT,\, 1}}{16\pi^2 (2+\eta_{4\rm F})}g_{{\rm GUT},\, \ast}^4.
\ee
This fixed point is infrared attractive, i.e., the value of the four-fermion-coupling at the Planck scale is predicted. At the same time, it serves as an upper bound for the asymptotically free case: If $g_{\rm GUT}$ is asymptotically free, the fixed-point value for the four-fermion coupling vanishes. While $g_{\rm GUT}$ increases from its asymptotically free fixed point towards its non-zero Planck-scale value, it pulls $G_{4\rm F}^{qqq\ell}$ along with it. To be concrete, we provide a numerical example. For $f_g=\frac{1}{4\pi}$, $\#_{\rm GUT,\, 1}=1$, $\#_{\rm GUT,\, 2}=40$ (translating into $g_{\rm GUT\, \ast}\approx 0.56$), cf.~\cite{Eichhorn:2017muy} and $\eta_{4F}\approx 0.1$, we obtain
\be
G_{4\rm F,\,\ast}^{qqq\ell}=- 3\cdot 10^{-4}.
\ee


\bibliography{references.bib}